\documentclass[aps,prl,twocolumn,superscriptaddress,showpacs]{revtex4-1}
\usepackage{amsmath,amssymb,wasysym,graphicx}

\usepackage[utf8]{inputenc}
\usepackage[T1]{fontenc}
\usepackage{xcolor}

\IfFileExists{newtxtext.sty}
   {\usepackage{newtxtext,newtxmath}}
   {\IfFileExists{stix.sty}
      {\usepackage{stix}}
      {\IfFileExists{mathptmx.sty}
      {\usepackage{mathptmx}}{} } }

\usepackage{textcomp}

\usepackage{bm}

\IfFileExists{siunitx.sty}{\usepackage{booktabs,siunitx}}{}

\pdfoutput=1
\usepackage{color}
\definecolor{LinkColor}{rgb}{0.256,0.439,0.588}
\usepackage{hyperref}
\hypersetup{
   pdfauthor={good guys},
   pdftitle={good title},
   colorlinks=true,
   citecolor=LinkColor,
   linkcolor=LinkColor,
   urlcolor=LinkColor
}

\renewcommand{\vec}[1]{\mathbf{#1}}

\usepackage{pifont}
\usepackage[normalem]{ulem}

\usepackage{multirow}

\begin{document}

\title{Valence Bond Orders at Charge Neutrality in a Possible Two-Orbital Extended Hubbard Model for Twisted Bilayer Graphene}

\author{Yuan Da Liao}
\affiliation{Beijing National Laboratory for Condensed Matter Physics and Institute of Physics, Chinese Academy of Sciences, Beijing 100190, China}
\affiliation{School of Physical Sciences, University of Chinese Academy of Sciences, Beijing 100190, China}
\author{Zi Yang Meng}
\affiliation{Department of Physics and HKU-UCAS Joint Institute of Theoretical and Computational Physics, The University of Hong Kong, Pokfulam Road, Hong Kong, China}
\affiliation{Beijing National Laboratory for Condensed Matter Physics and Institute of Physics, Chinese Academy of Sciences, Beijing 100190, China}
\affiliation{Songshan Lake Materials Laboratory, Dongguan, Guangdong 523808, China}
\affiliation{CAS Center of Excellence in Topological Quantum Computation and School of Physical Sciences, University of Chinese Academy of Sciences, Beijing 100190, China}
\author{Xiao Yan Xu}
\email{wanderxu@gmail.com}
\affiliation{Department of Physics, Hong Kong University of Science and Technology, Clear Water Bay, Hong Kong, China}
\affiliation{Department of Physics, University of California at San Diego, La Jolla, California 92093, USA}

\date{\today}

\begin{abstract}
An extended Hubbard model on a honeycomb lattice with two orbitals per site at charge neutrality is investigated with unbiased large-scale quantum Monte Carlo simulations. The Fermi velocity of the Dirac fermions is renormalized as the cluster charge interaction increases, until a mass term emerges and a quantum phase transition from Dirac semi-metal to valence bond solid (VBS) insulator is established. 
The quantum critical point is discovered to belong to 3D $N=4$ Gross-Neveu chiral XY universality with the critical exponents obtained at high precision. Further enhancement of the interaction drives the system into two different VBS phases, the properties and transition between them are also revealed. Since the model is related to magic-angle twisted bilayer graphene, our results may have relevance towards the symmetry breaking order at the charge neutrality point of the material, and associate the wide range of universal strange metal behavior around it with quantum critical fluctuations.
\end{abstract}

\maketitle

{\it Introduction}\,---\,
Twisted bilayer graphene (TBG) forms Mori\'e patterns in real space with the size of the Mori\'e unit cell tuned by the twisting angle. The Fermi velocity of the Dirac fermions of monolayer graphene is renormalized in TBG. At some magic angle, the Fermi velocity vanishes~\cite{bistritzer2011moire,morell2010flat,santos2012continuum,fang2016electronic,trambly2012numerical,Tranopolsky2018}, such that flat bands are formed and the system consequently become susceptible towards many instabilities. In 2018, the gate tunable magic-angle TBG is realized in laboratory~\cite{cao2018correlated,cao2018unconventional}, and interesting phenomena encompassing the correlated insulating phase~\cite{cao2018correlated}, unconventional superconductivity~\cite{cao2018unconventional,Yankowitzeaav2019} and strange metal behavior~\cite{YuanCao2019} are quickly discovered. Those results hint that, unlike its monolayer cousin, the gate tunable magic-angle TBG is a strongly correlated system in nature and share many common features of the phase diagram of doped cuprates, consequently spur the interests of theoretical and experimental communities on Mori\'e physics~\cite{chen2018gate,xu2018topo,yuan2018model,Po2018PRX,Po2018,liu2018chiral,dodaro2018phases,huang2018antiferromagnetically,roy2019,guo2018pairing,xu2018kekule,zhang2018nearly,koshino2018,wu2018theory,tang2018,YuanCao2019,Yankowitzeaav2019,Zou2018,Kang2018,thomson2018,venderbos2018,zhu2018inter,zhang2018low,you2018superconductivity,lian2018twisted,song2018all,liu2018complete,xie2018nature,zhu2018spin,ChengShen2019,XiaomengLiu2019,lu2019}.
 
Compared with cuprates, magic-angle TBG also  acquires unique properties and two of them are related with the modeling of the material. First, although there are huge number of electrons in one unit cell which fill thousands of energy bands, various band calculations show that there exist an isolated band branch with four bands around charge neutrality point~\cite{bistritzer2011moire,morell2010flat,santos2012continuum,fang2016electronic,trambly2012numerical,koshino2018,Kang2018}.  The four bands are made up of the spin and valley degrees of freedom of untwisted graphene. Second, the charge center forms a triangular lattice, but symmetry obstacles force one to define the effective model on a honeycomb lattice if the different band degeneracy at $\Gamma$ and $\mathbf{K}$ of the BZ were to be respected~\cite{yuan2018model,Po2018PRX}. Meanwhile, there are also obstacles from deriving a single valley tight-binding model due to chirality or mirror symmetry~\cite{Zou2018}. Putting all these factors together, a two orbital (counts the two valleys of the untwisted graphene) spinful lattice model on a honeycomb lattice with cluster charge interaction (considering the charge center form triangle lattice) is a good starting point to describe the system~\cite{Po2018PRX,koshino2018,xu2018kekule,kang2019}. 

But such a model is still a strongly correlated one and cannot be solved analytically. In light of the situation, we performed unbiased sign-problem-free quantum Monte Carlo (QMC) simulation to investigate such a system at charge neutrality and map out its precise phase diagram. By gradually increasing the interaction strength, the phase diagram exhibits -- in a consecutive manner -- a Dirac semi-metal (DSM), a plaquette valence bond solid (pVBS) and a columnar valence bond solid (cVBS) phases at weak, intermediate and strong interaction regions. The quantum phase transitions between these phases are revealed with scrutiny and we found the DSM-pVBS transition is continuous, belonging to the 3D $N=4$ Gross-Neveu chiral XY universality class; the pVBS-cVBS transition on the other hand is first order but bestowed with a sign change in the mass term of fermion bilinear, and implies that a quantum pseudo-spin Hall effect can be generated between the zigzag domains of those two insulators. The experimental relevance of our discoveries in quantum criticality and phase transitions towards to on-going investigations of TBGs is also discussed.

{\it Model and Method}\,---\,
We study an extended Hubbard model with two orbitals of spinful fermions on a honeycomb lattice. The model contains two parts $H=H_{0}+H_U$, where
\begin{equation}
\label{eq:tb}
H_{0} =  -t\sum_{\langle ij \rangle l \sigma}\left(c^{\dagger}_{il\sigma}c_{jl\sigma}+h.c. \right)-t_2\sum_{\langle ij \rangle' l\sigma} \left( i^{2l-1}c^{\dagger}_{il\sigma}c_{jl\sigma} + h.c. \right)
\end{equation}
is the tight-binding part introduced in Ref.~\cite{koshino2018} and serves as a minimal model to describe of the low energy band structure of magic-angle TBG with Dirac points at charge neutrality and band splitting along $\Gamma$-M direction.
% (see Fig.S1 in Sec.I of Supplemental Material (SM)~\cite{suppl}).
Here $c^{\dagger}_{il\sigma}$ ($c_{il\sigma}$) is the creation (annihilation) operator of electron at site $i$, orbital $l=1,2$ with spin $\sigma=\uparrow,\downarrow$. Throughout this Letter, we take the nearest neighbor hopping $t$ as the energy unit. The fifth neighbor hopping ($it_2$ for $l=1$ and $-it_2$ for $l=2$) is purely imaginary and breaks orbital degeneracy along $\Gamma$-M direction.
%, also shown in the Fig.S1 of SM~\cite{suppl}. 
As $t_2/t$ is small in the material, we focus on the range of $t_2/t<0.6$. 

\begin{figure}[t!]
\includegraphics[width=\columnwidth]{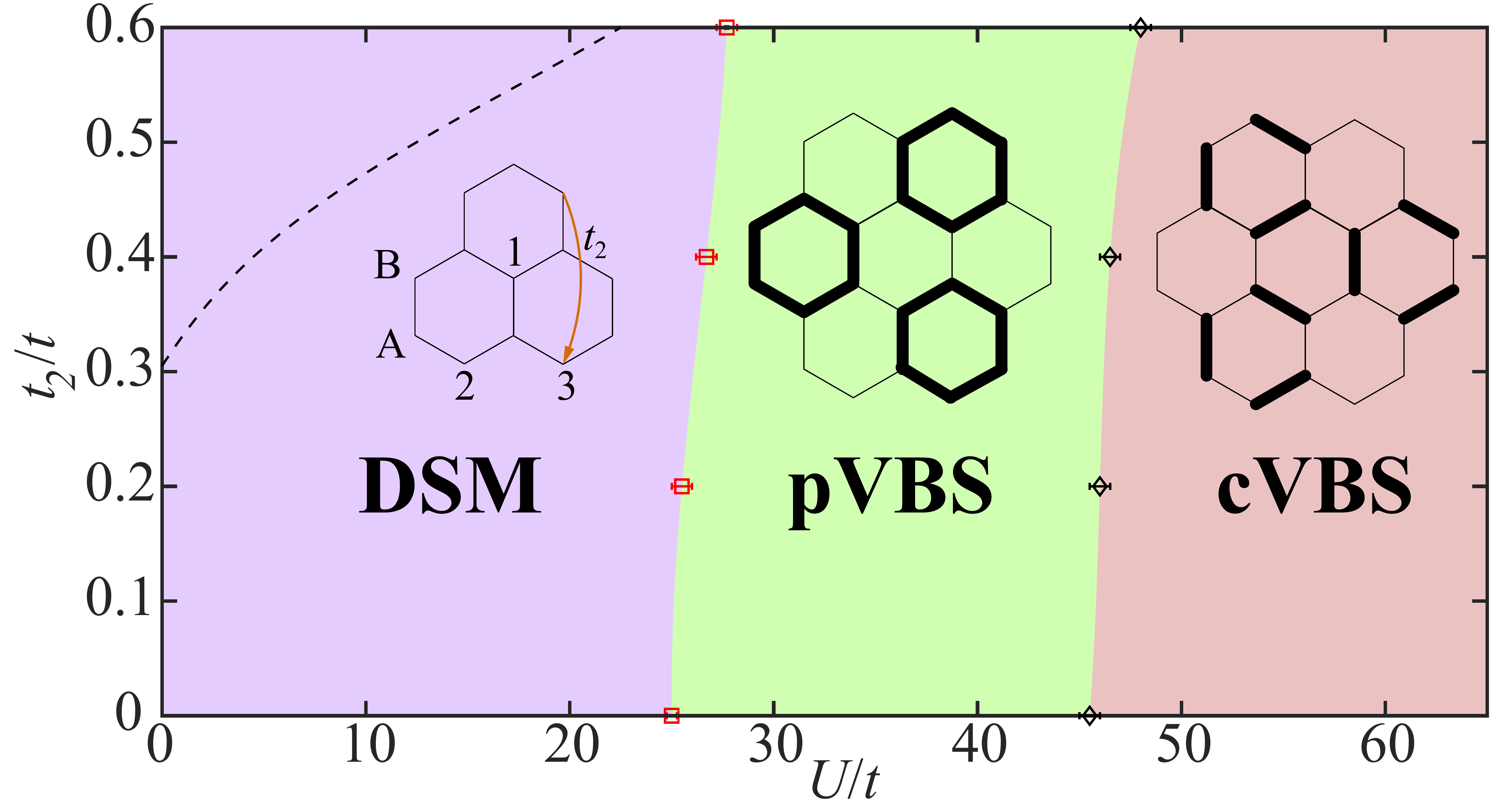}
\caption{Ground state phase diagram. As a function of $U/t$, DSM, pVBS and cVBS phases reveal themselves. The fifth neighbor hopping strength $t_2$ is marked as orange line in the inset. Inside DSM, the $t_2/t$ ratio modifies the band degeneracy and Fermi surface topology, and the black dash line signifies the corresponding band structure crossover, where excitons formed between orbitals might condense. The different patterns of the valence bonds are shown in the insets. The DSM-pVBS transition is continuous, and shown to belong to 3D $N=4$ chiral XY Gross-Neveu universality. The pVBS-cVBS transition is first order, but could carry topological edge state in terms of quantum pseudo-spin Hall effect in the pVBS-cVBS zigzag domain wall.}
\label{fig:phasediagram}
\end{figure}

For the Coulomb interaction term $H_U$, as the Wannier orbitals are quite extended in TBG, onsite, first, second and third neighbor repulsions are all important~\cite{koshino2018,Po2018PRX,kang2019}. To capture such non-local interactions, a cluster charge Hubbard term which maintain the average filling of each elemental hexagon on the honeycomb lattice to be 4 is the genuine choice, therefore we write down
\begin{equation}
\label{eq:interaction}
H_U = U\sum_{\varhexagon}(Q_{\varhexagon}-4)^2,
\end{equation}
where the cluster charge $Q_{\varhexagon} \equiv \sum_{i\in \varhexagon}\frac{n_i}{3}$ with $n_i = \sum_{l\sigma}c^\dagger_{il\sigma}c_{il\sigma}$ summing over all the six sites of the elemental hexagon. If we expand Eq.~\eqref{eq:interaction}, the  onsite, first, second and third neighbor interaction strength are $\frac{2}{3}U$, $\frac{4}{9}U$, $\frac{2}{9}U$ and $\frac 2 9 U$, with ratio 3:2:1:1. As different range of interactions favor different kinds of order, it may require larger interaction to open a fermion gap compared to a local Hubbard model.

At any finite $U/t$, the model $H=H_{0}+H_{U}$ is nonperturbative in nature, but we found it actually immune from sign-problem due to an antiunitary symmetry~\cite{wu2005suff} at charge neutrality point, 
% as detailed in Sec.II of SM~\cite{suppl},
and is readily exposed to large-scale projection QMC (PQMC) simulations~\cite{blankenbecler1981monte,hirsch1985two,Assaad2008}. PQMC speaks out the ground state phase diagram, correlation functions (to determine the pattern of symmetry-breaking) and dynamical information (single-particle and collective excitation gaps above the ground state), and has been employed in several our previous studies~\cite{meng2010quantum,xu2017topo,YuanYaoHe2018,xu2018kekule}.  The symmetry analysis and numeric implementation of PQMC are discussed in SM~\cite{suppl}, we only mention here that the projection length is set to $\Theta = 2L$ and the simulations are performed with linear system size upto $L=24$, amounts to $N_e=4\times L^2=2304$ interacting electrons on TBG model.

{\it Phase diagram and quantum criticality}\,---\,  
Our phase diagram, expanded by axes $U/t$ and $t_{2}/t$, is shown in Fig.~\ref{fig:phasediagram}, the DSM, pVBS and cVBS phases are in place. The two VBS are gapped insulators. In the following parts, $t_2/t=0$ if it is not specified. It is interesting to notice that different from the local Hubbard~\cite{sorella1992,meng2010quantum}, $t-J$~\cite{Lang2013} and extended cluster charge models with single orbital~\cite{xu2018kekule} on honeycomb lattice, the AB sublattice antiferromagnetic insulating phase is suppressed in our phase diagram even at very large $U/t$, which may be understood by a perturbation theory in the large $U/t$ limit~\cite{zhou2016mott}, but the existence of two VBS phases in such a simple model is unexpected and has not been found in a local Hubbard model before.
 
\begin{figure}[t!]
\includegraphics[width=\columnwidth]{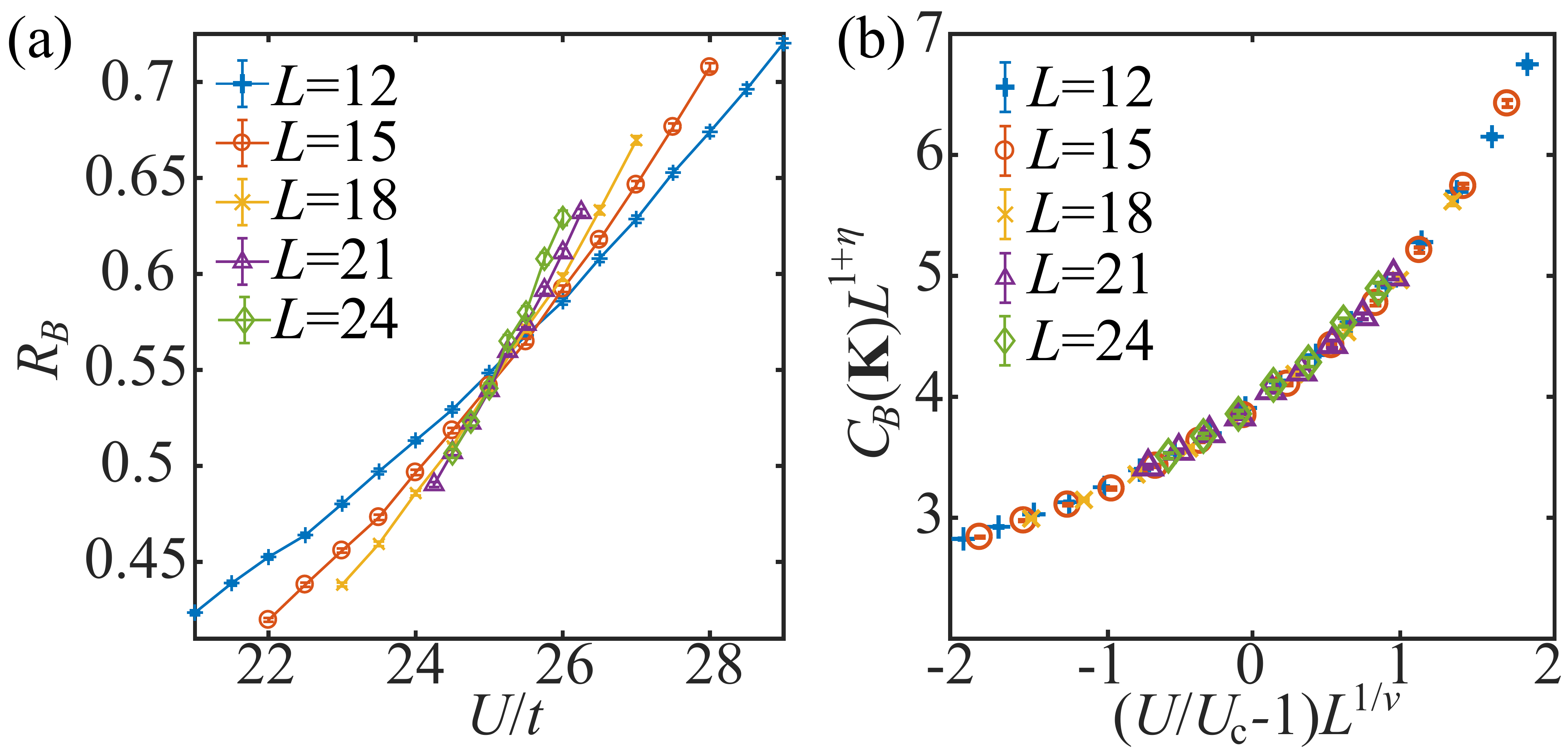}
\caption{(a) The bond-bond correlation ratio $R_{B}$ and (b) data collapse analysis of structure factor $C_{B}(\mathbf{K})$ at $t_2/t=0$ as function of $U/t$ with $L=12,15,\cdots,24$. The crossing of $R_{B}$ in (a) gives the DSM-pVBS critical point $U_c/t = 25.1(2)$. The data collapse in (b) gives the 3D $N=4$ Gross-Neveu chiral XY exponents $\eta=0.80(2)$, $\nu=1.01(3)$.}
\label{fig:crossandcollapse}
\end{figure}

To study the DSM-pVBS transition, we measure the bond-bond structure factor,
\begin{equation}
C_B(\mathbf{k}) = \frac{1}{L^4}\sum_{i,j}e^{i\vec{k}\cdot(\mathbf{r}_i - \mathbf{r}_j)}\left\langle B_{i,\delta} B_{j,\delta} \right\rangle
\end{equation} 
where bond operator $B_{i,\delta}=\sum_{l,\alpha} (c_{i,l,\alpha}^\dagger c_{i+\delta,l,\alpha}+h.c.)$ with $\delta$ standing for one of the three nearest-neighbor bond directions ($\hat{e_1}$, $\hat{e}_2$ and $\hat{e}_3$) and $\hat{e}_1$ is chosen in the calculation. Results show that $C_{B}(\mathbf{k})$ is peaked at momenta $\mathbf{K}$ and $\mathbf{K'}$ $(\pm \frac{4\pi}{3\sqrt{3}a_0}, 0)$ of the BZ suggesting the VBS patterns shown in the inset of Fig.~\ref{fig:phasediagram}. 

To locate the DSM-pVBS transition point, we plot the correlation ratio $R_B(U,L)=1-\frac{C_B(\mathbf{K}+\delta \mathbf{q})}{C_B(\mathbf{K})}$ for different $U$ and system size $L$ with $|\delta\mathbf{q}|\sim\frac{1}{L}$. This quantity approaches to one (zero) in an ordered (disordered) phase, and implies a crossing for different $L$ at an critical point~\cite{kaul2015} as showed in Fig.~\ref{fig:crossandcollapse}(a), where one reads $U_c/t=25.1(2)$. 
We further collapse the bond-bond structure factor with scaling relation $C_B(\mathbf{K},U,L)=L^{-(1+\eta)}f(L^{1/\nu}(U-U_c)/U_c)$  
(dynamical exponent $z$ is set to 1 due to Lorentz symmetry of massless Dirac fermions),
as shown in Fig.~\ref{fig:crossandcollapse} (b). 
and obtain the critical exponents $\eta = 0.80(2)$ and $\nu = 1.01(3)$. Since our Dirac fermions acquire 4 degrees of freedom per site and the pVBS phase contains an emergent $U(1)$ symmetry close to the DSM-pVBS transition as shown in Refs.~\cite{xu2018kekule,zhou2016mott}, we identify this transition in the 3D $N=4$ Gross-Nevue chiral XY universality class~\cite{gross1974,hands1993,rosenstein1993critical,Zerf2017,li2017fermion,zhou2016mott,scherer2016gauge,mihaila2017gross,jian2017fermion,classen2017fluctuation,torres2018fermion,Ihrig2018}. The critical exponents obtained here ($\eta=0.80(2),\nu=1.01(3)$) are comparable with those calculated theoretically or numerically in literatures, as showed in Table.I of SM~\cite{suppl}.

\begin{figure}[t!]
\includegraphics[width=\columnwidth]{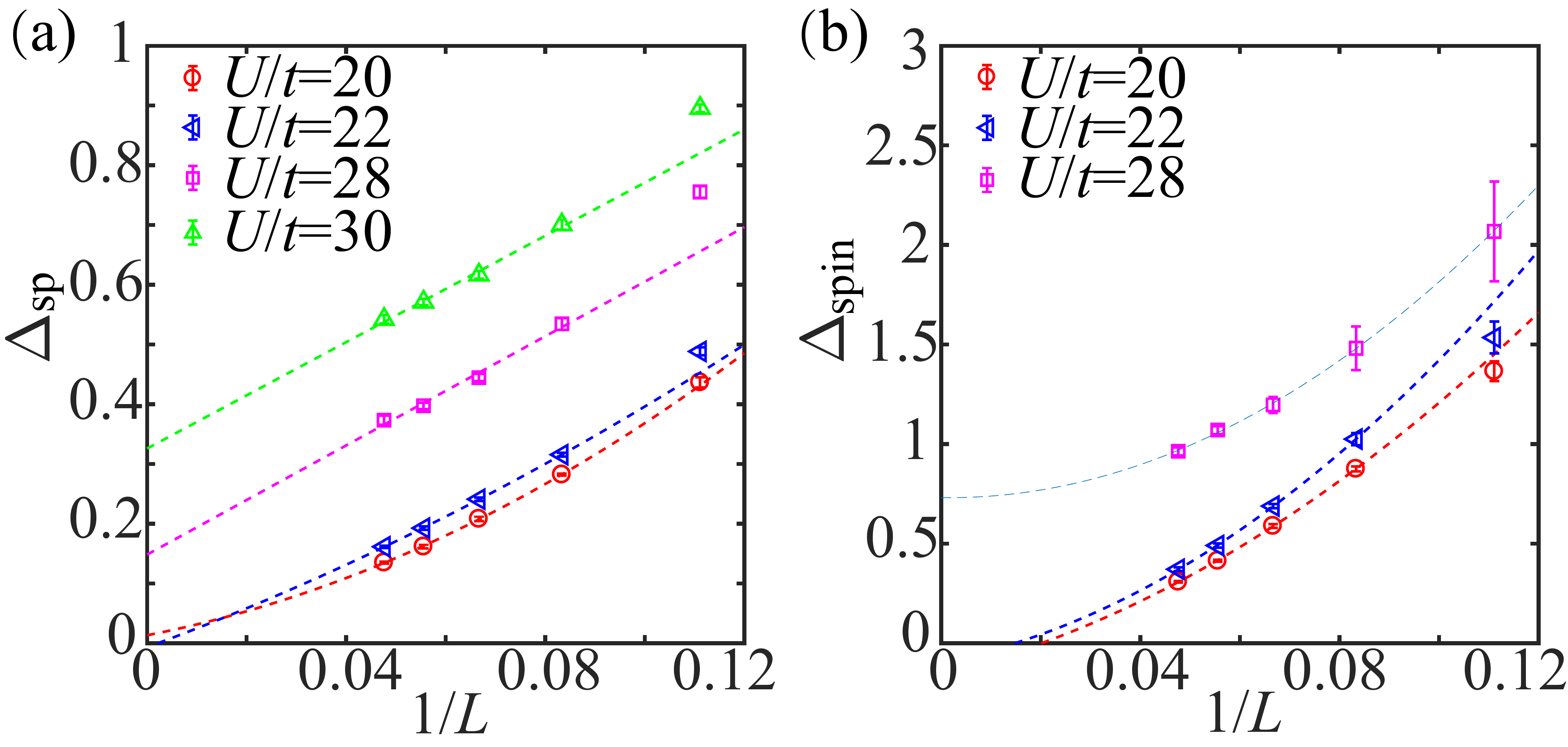}
\caption{(a) The $1/L$ extrapolation of single-particle gap $\Delta_{\text{sp}}(\mathbf{K})$, the gap opens between $U/t=22$ and $U/t=28$, consistent with the $U_c/t$ obtained from the bond correlation ratio in Fig.~\ref{fig:crossandcollapse} (a). (b) The $1/L$ extrapolation of spin gap $\Delta_{\text{spin}}(\mathbf{K})$, the spin gap opens hand-in-hand with the single-particle gap as the establishment of pVBS order.}
\label{fig:gap}
\end{figure}

{\it Gapped phases: cVBS and pVBS}\,---\, 
The DSM is known to possess robust massless linear dispersion at weak interaction ($U<U_c$)~\cite{sorella1992,assaad2013pinning,otsuka2016universal,zhou2016mott,sato2017dirac,
xu2018kekule,ChuangChen2018}, and the Dirac fermion will be gapped out in the pVBS insulator. To monitor the opening of the single-particle gap across the DSM-pVBS transition, we measure the dynamical single-particle Green's function and follow its decay in imaginary time $G(\mathbf{k},\tau) \propto e^{-\Delta_{\text{sp}}(\mathbf{k})\tau}$ at momentum $\mathbf{K}$ for increasing system size $L$ and imaginary time displacement $\tau$, with $G(\mathbf{k},\tau)=\frac{1}{4L^2}\sum_{i,j,l,\sigma}e^{i\mathbf{k}\cdot (\mathbf{r}_i-\mathbf{r}_j)}\langle  c_{i,l,\sigma}(\frac{\tau}{2}) c^{\dagger}_{j,l,\sigma}(-\frac{\tau}{2})\rangle$. The obtained $\Delta_{\text{sp}}$ for different interaction $U$ and $L$ are shown in Fig.~\ref{fig:gap} (a). It is clear that when $U<U_c$, $\Delta_{\text{sp}}\to 0$ and when $U>U_c$, $\Delta_{\text{sp}}$ goes to a finite value, which validate the picture that the DSM-pVBS transition is accompanied by the opening of the single-particle gap.

\begin{figure}[h!]
\includegraphics[width=\columnwidth]{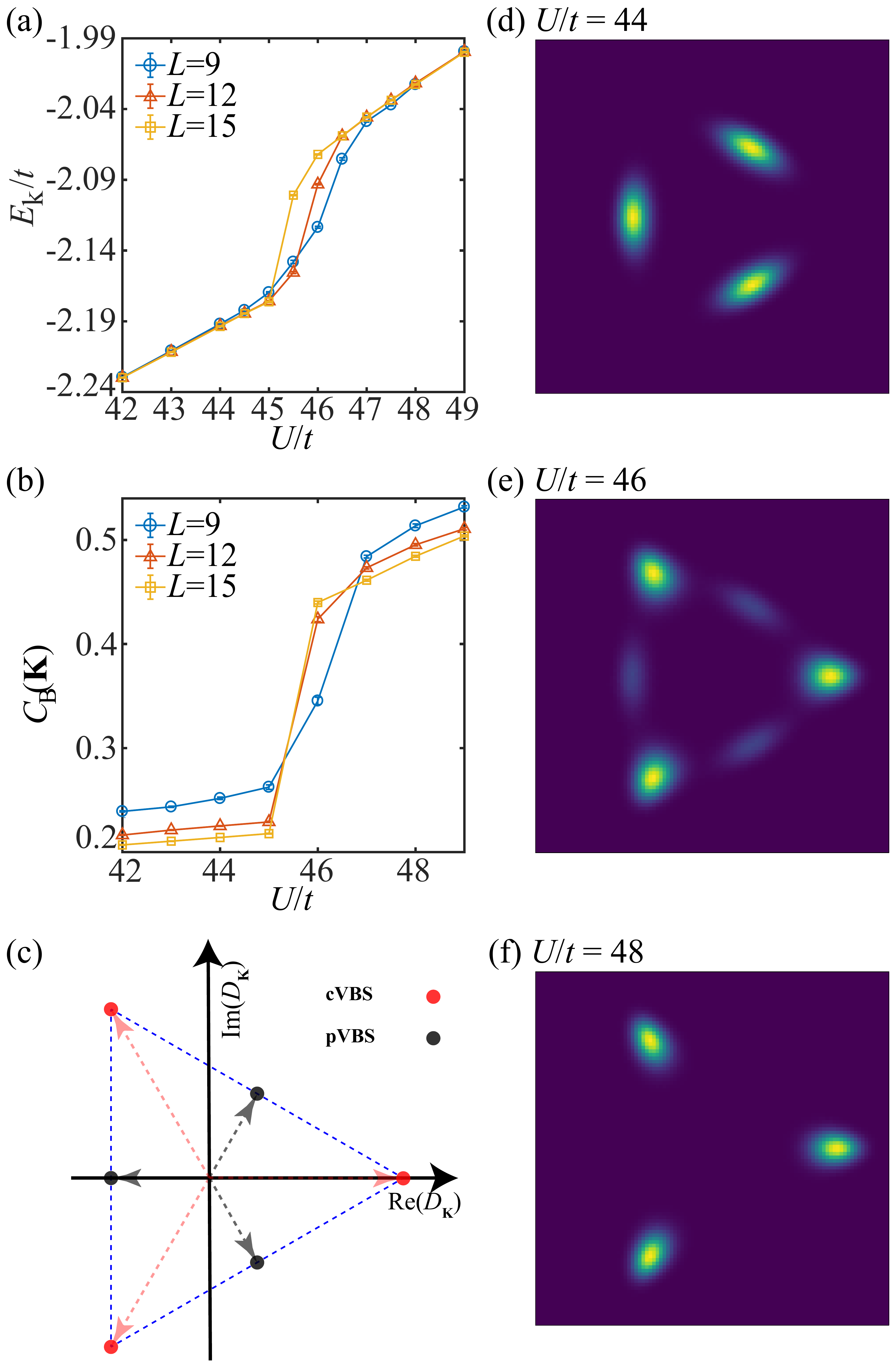}
\caption{(a) Kinetic energy per site of the system for $U$ at large values. The sharp jump signifies a first order transition. (b) $C_{B}(\mathbf{K})$ for the same process, a jump in VBS order is also observed, suggesting this is a transition between different VBS phases. (c) Angular dependence of the complex order parameter $D_\mathbf{K}$. Black dots represent ideal pVBS order, and red dots represent ideal cVBS order. (d)-(e) Histogram of $D_\mathbf{K}$ at different interaction strengths $U<U_{\text{VBS}}$, $U\approx U_{\text{VBS}}$ and $U>U_{\text{VBS}}$.}
\label{fig:secondPT}
\end{figure}

The VBS phase is related with the formation of spin singlet, either within the hexagon plaquette (pVBS) or along the nearest-neighbor bond (cVBS), and to see that one can examine the spin excitation gap, obtained from the imaginary time decay of dynamical spin-spin correlation function $C_S(\mathbf{q},\tau) = \frac{1}{N}\sum_{i,j} e^{i\mathbf{q}\cdot (\mathbf{r}_i-\mathbf{r}_j)} \langle \mathbf{S}_{i}(\frac {\tau} 2)\mathbf{S}_{j}(-\frac {\tau}{2})\rangle$ as $C_S(\mathbf{K},\tau) \propto e^{-\Delta_{\text{spin}}(\mathbf{K})\tau}$. The spin operator is defined as $\mathbf{S}_{i}=\frac 1 2 \sum_{l,\alpha,\beta}c^{\dagger}_{i,l,\alpha}(\boldsymbol{\sigma})_{\alpha,\beta}c_{i,l,\beta}$ for $t_2/t\ne0$ case with $\boldsymbol{\sigma}=(\sigma^1, \sigma^2, \sigma^3)$, and $(\vec{S}_i)_\mu ^\nu = c_{i,\mu}^\dagger c_{i,\nu} - \frac{\delta_{\mu\nu}}{4} \sum_{\rho=1}^{4} c_{i,\rho}^\dagger c_{i,\rho} $ for $t_2/t=0$ case where $\mu$, $\nu$ and $\rho$ denote combination of indexes of spin and orbital as components of SU(4) generator. Similar to the case of the single-particle gap, after extrapolation of $\Delta_{\text{spin}}(\mathbf{K})$ ($\Delta_{\text{spin}}$ is the smallest and degenerate at momenta $\mathbf{K}$, $\mathbf{K'}$ and $\Gamma$) for various $L$ and $U$, one can see from Fig.~\ref{fig:gap} (b) that the spin gap is also zero when $U<U_c$ in the DSM phase and becomes finite when $U>U_c$ as the system enters the pVBS phase.

Further increase $U/t$ from the pVBS phase, a kinetic energy jump at $U/t\approx 46$ is observed, as shown in Fig.~\ref{fig:secondPT} (a). At the same time, the VBS correlation $C_B(\mathbf{K})$ also acquires a jump at the same $U$ as shown in Fig.~\ref{fig:secondPT} (b). These results point out that, besides $U_c/t=25.1(2)$ there is another first order phase transition at $U_{\text{VBS}}/t \approx 46$ between two different VBS phases. There are three non-equivalent VBS configurations, but only two of them, the pVBS and cVBS as depicted in the inset of Fig.\ref{fig:phasediagram}, can perturbatively open the single-particle gap to VBS phases. And the jumps observed in Fig.~\ref{fig:secondPT} (a) and (b), might be the transition between these two VBS phases.

To verify this idea, we follow Refs.~\cite{Lang2013,zhou2016mott} and make use of the nearest-neighboring bonds $B_{i,\delta}$ originated from sublattice A to construct a complex order parameters $D_{\mathbf{K}} = \frac{1}{L^2}\sum_{i}\left( B_{i,\hat{e}_1}+\omega B_{i,\hat{e}_2} +\omega^2 B_{i,\hat{e}_3}\right) e^{i\mathbf{K}\cdot\mathbf{r}_i}$ with $\omega = e^{i\frac{2\pi}{3}}$. The Monte Carlo histogram of $D_{\mathbf{K}}$ can reveal the difference between the two VBS phases. As shown in Fig.~\ref{fig:secondPT} (c). The angular distribution of pVBS will peak at ${\rm arg} (D_{\mathbf{K}}) = \frac{\pi}{3}, \pi, \frac{5\pi}{3}$, whereas that of cVBS will peak at ${\rm arg} (D_{\mathbf{K}}) = 0,\frac{2\pi}{3},\frac{4\pi}{3}$. Fig.~\ref{fig:secondPT} (d), (e) and (f) show the corresponding histograms at three representative interaction strengths $U=44t < U_{\text{VBS}}$, $U=46t \approx U_{\text{VBS}} $, $U=48t > U_{\text{VBS}}$. It is clear that Fig.~\ref{fig:secondPT} (d) and (f) are inside pVBS and cVBS, respectively, Fig.~\ref{fig:secondPT}(e) on the other hand depicts the distribution of both characters, a typical example of the coexistence at the first order transition point. In this way, the two VBS phases and their first order transition are clearly established. Similar scenario, between two Kekul\'e patterned superconducting states, in the context of attractive interaction on honeycomb lattice has been discussed in Ref.~\cite{Roy2010}.

\begin{figure}[t!]
\includegraphics[width=1.0\columnwidth]{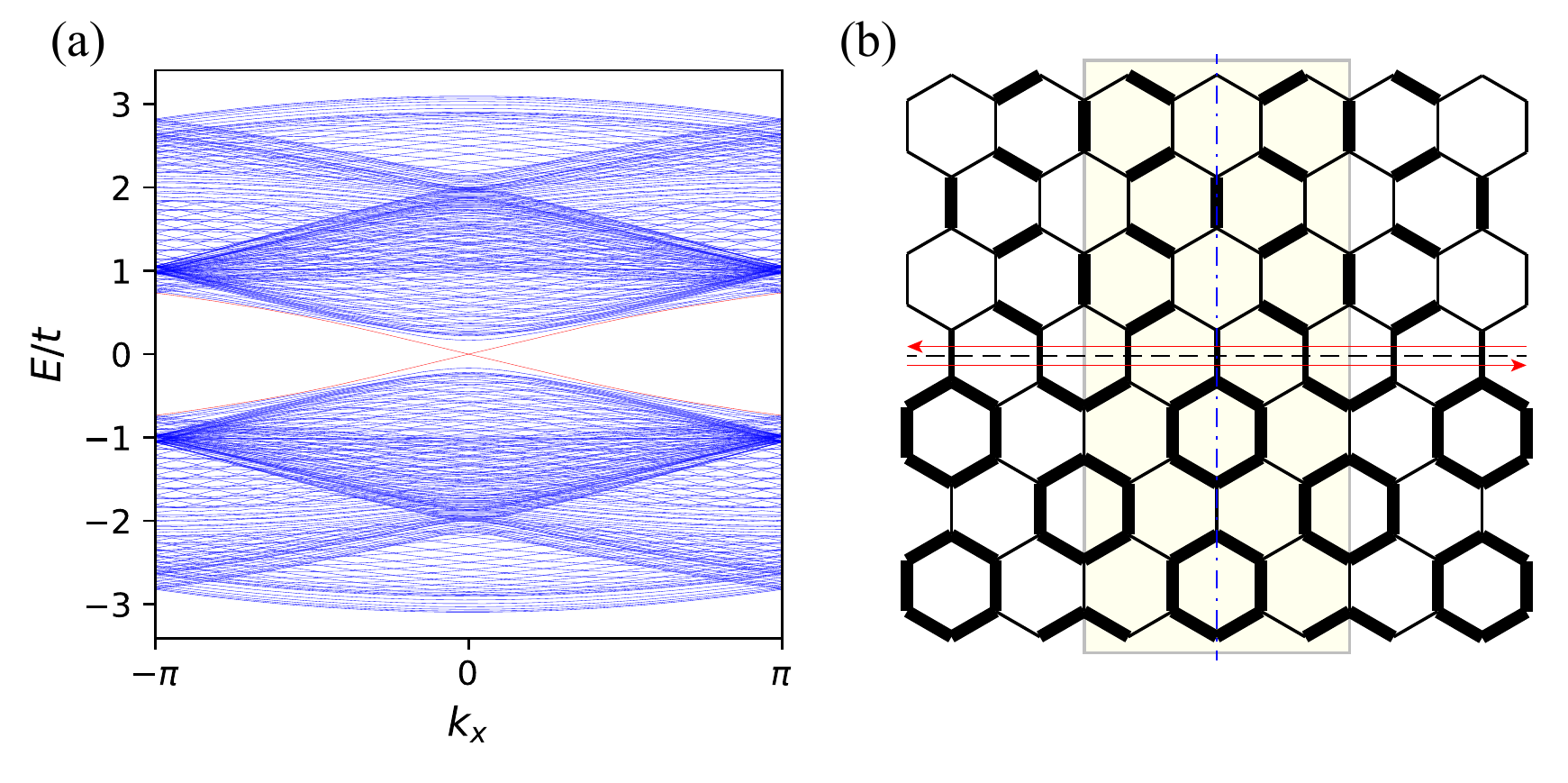}
\caption{(a) Spectrum of zigzag domain wall of pVBS and cVBS. The red line denotes the helical edge states. In the calculation, both the width of pVBS and cVBS strips are set to 24 times of honeycomb unit cell. (b) The zigzag domain wall of pVBS and cVBS. Bottom part is pVBS phase and top part is cVBS phase. The hopping strength is $-1.1t$ for strong bonds and $-0.9t$ for weak bonds. The pVBS and cVBS are connected by vertical bonds with hopping strength $-t$.  The domain wall has periodic boundary condition along $x$ direction, and is put on a torus (upper boundary and lower boundary are also connected by vertical bonds with hopping strength $-t$). The shaded region denotes the super unit cell of the domain wall.
}
\label{fig:domain}
\end{figure}

{\it Pseudo spin Hall effect in zigzag domain of pVBS and cVBS}\,---\, 
Our model provides the unique opportunity that pVBS and cVBS  all appear in the phase diagram due to spontaneous Dirac mass generation, and one can reveal their connection with the following analysis. In the VBS phase we consider a mean field description with bond charge order, and  suppress the spin and orbital degrees of freedom for the moment. 
Then the static VBS order becomes a modulation in the nearest neighbor hopping. As showed in the inset of Fig.~\ref{fig:phasediagram}, there are two kinds of bonds and the hopping magnitude is defined as $(1+\delta)t$ and $(1-\delta)t$. From a tight binding Hamiltonian with such bond modulation, a $4\times4$ $k\cdot p$ Hamiltonian at $\Gamma$ point can be derived,
\begin{equation}
\label{eq:kp}
H_{\text{eff}}(\vec{k}) = -t\left( \tilde{\vec{k}}\cdot \tilde{ \vec{s}}\tau^2 + ms^0 \tau^3 \right),
\end{equation}
where momentum $\tilde{\vec{k}}\equiv (\frac 3 2 k_y, \frac {\sqrt{3}} {2} ik_x, \sqrt{3}k_x)$, the vector $\tilde{\vec{s}}=(s^1, s^2, s^3)$ and mass term $m=2\delta$. Here $s^i$ and $\tau^i$ are Pauli matrices in two different spaces. From the $k\cdot p$ Hamiltonian, it is clear that the bond modulation plays the role of a mass term, and there is a sign change in it across the pVBS-cVBS transition.

The sign change in the mass term motivates us to study the domain walls between pVBS and cVBS phases. Both VBS phases coexist at the first order transition point and may go through a gap close at the domain wall. Interestingly, the calculation of the spectrum of the pVBS and cVBS zigzag domain wall on a torus shows that, there exist robust helical edge states as depicted in the Fig.~\ref{fig:domain}. Such edge states are protected by the combined symmetry of sublattice (chiral) and mirror (along a bond with mirror plane denoted as blue dashed line in Fig.~\ref{fig:domain}(b)) and can be interpreted as the edge states of quantum pseudo spin Hall effect~\cite{kariyado2017,liu2017pseu,sorella2018structural}.

{\it Experimental relevance}\,---\, 
Recent STM experiments~\cite{kerelsky2018,choi2019} find significant gap opening near charge neutrality point (CNP) indicating symmetry breaking order. The insulating phase at CNP is also found in many other experiments~\cite{Yankowitzeaav2019,YuanCao2019,lu2019}. The spontaneous symmetry breaking VBS phases found in our simulations hence provide good candidates, and a Fourier transform of the large-area STM topograph may detect addtional features at the wavelength of VBS~\cite{gutierrez2016imaging}. Although exact estimation of cluster charge interaction $U/t$ is not easy, our results hint the possibility that magic-angle TBG is close to a QCP where the Dirac mass is spontaneously generated and opens the charge neutrality gap. Other exciting observation shows a wide range of strange metal behavior, whose existence seems robust against experimental details~\cite{YuanCao2019}. Such universal transport behavior is the hallmark of quantum critical phenomena, possibly originated from the 3D Gross-Neveu chiral XY transition between DSM and pVBS discovered here. Moreover, the strange metal behavior is pronounced near $\pm 1/4$ filling, and not well established near charge neutrality point, this is consistent with the picture that the charge neutrality point is gapped but close to the QCP, when gated, the quantum critical fluctuation kicks in and generates strange metal behavior. A final remark is that the model we studied assumed well defined valley degrees of freedom (in terms of two orbitals). Actually, if the valleys coupling is considered in TBG, it will give a single orbital model~\cite{Po2018PRX}. The interesting thing is that we still find a QCP at moderate interaction strength in the single orbital model which may be related with the intriging physics near $\pm 1/4$ filling of magic-angle TBG~\cite{xu2018kekule}.

\begin{acknowledgments}
{\it Acknowledgments}\,---\, 
We thank Noah Yuan, Liang Fu, Eslam Khalaf,  Lukas Janssen, Yang Qi, Qing-Rui Wang, Chen Fang, K. T. Law and Patrick Lee for helpful discussions. We thank Michael Scherer for providing us the $4-\epsilon$ four loop calculation of critical exponents for 3D $N=4$ Gross-Neveu chiral XY universality class. Y. D. L. and Z. Y. M. acknowledge support from the Ministry of Science and Technology of China
through the National Key Research and Development
Program (2016YFA0300502), the Strategic Priority
Research Program of the Chinese Academy of Sciences
(XDB28000000), the National Science Foundation of
China (11574359) and Research Grants Council of Hong
Kong Special Administrative Region of China through
17303019. X. Y. X. is thankful for the support of Research
Grants Council of Hong Kong Special Administrative
Region of China through C6026-16W. We thank the Center for Quantum Simulation Sciences at Institute of Physics,  Chinese Academy of Sciences, and the Tianhe-1A platform at the National Supercomputer Center in Tianjin for technical support and generous allocation of CPU time.
\end{acknowledgments}

\bibliographystyle{apsrev4-1}
\bibliography{main}

\begin{thebibliography}{76}
\bibitem [{sup()}]{suppl}%
  \BibitemOpen
  \href@noop {} {\bibinfo  {journal} {Tight-binding model of the TBG, its
  dispersion and the Fermiology at various $t_2/t$, the symmetry analysis of
  the model and absence of sign-problem and its PQMC implementations,
  derivation of $k\cdot p$ Hamiltonian for VBS phase and the critical
  properties of the $N=4$ Gross-Neveu chrial XY transition are discussed in
  detail}\ }\BibitemShut {NoStop}%
\end{thebibliography}

\clearpage
\onecolumngrid
%\appendix
\begin{center}
\textbf{Supplemental Material for "Valence Bond Orders at Charge Neutrality in a Possible Two-Orbital Extended Hubbard Model for Twisted Bilayer Graphene"}
\end{center}
\setcounter{equation}{0}
\setcounter{figure}{0}
\setcounter{table}{0}
\setcounter{page}{1}
\makeatletter
\renewcommand{\thetable}{S\arabic{table}}
\renewcommand{\theequation}{S\arabic{equation}}
\renewcommand{\thefigure}{S\arabic{figure}}
\setcounter{secnumdepth}{3}

\section{Symmetries and band structure for different $t_2/t$}
The tight binding part of our Hamiltonian has the following form
\begin{equation}
\label{Seq:tb}
H_{0} =  -t\sum_{\langle ij \rangle l \sigma}\left(c^{\dagger}_{il\sigma}c_{jl\sigma}+h.c. \right)-t_2\sum_{\langle ij \rangle' l\sigma} \left( i^{2l-1}c^{\dagger}_{il\sigma}c_{jl\sigma} + h.c. \right).
\end{equation}
on a honeycomb lattice, where 
$t$ is the hopping amplitude between the nearest neighbor sites $\langle \cdots \rangle$, $t_2$ is the hopping amplitude between the fifth neighbor sites $\langle \cdots \rangle '$ and $i^{2l-1}$ sets the correct phase factor for orbital $l$, according to Ref.~\cite{koshino2018}. As there are two orbitals $l=1,2$ and two sites in one unit cell, there are four bands (do not count spin) in the Brillouin zone (BZ).

\begin{figure}[h!]
\includegraphics[width=0.4\columnwidth]{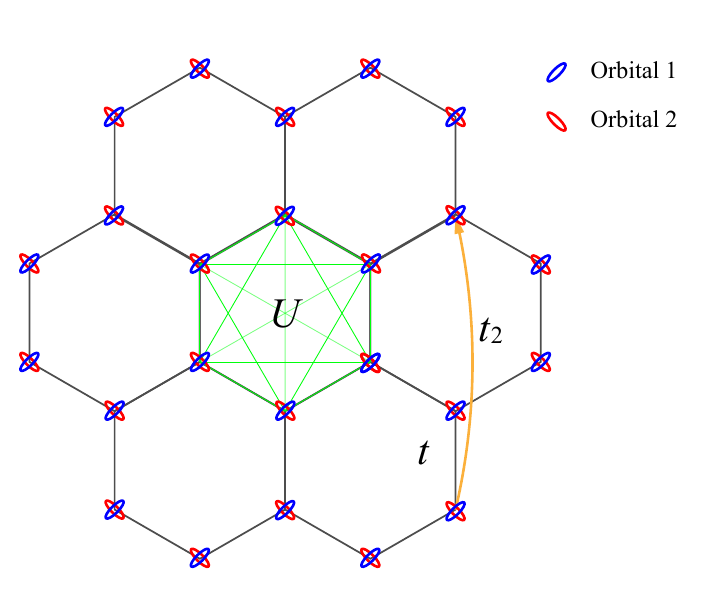}
\caption{ A two-orbital extended Hubbard model on a honeycomb lattice. $t$ is the nearest neighbor hopping, $t_2$ is the fifth neighbor hopping, $U$ is the cluster charge interaction.}
\label{Sfig:model}
\end{figure}

For generic $t_2/t$, $H_0$ acquires the symmetry group $G=D_3 \times U(1) \times SU(2) \times T \times \mathcal{Z}_2$, where $U(1)$ stems from charge conservation of each orbital, $SU(2)$ is the spin rotational symmetry, $T$ is time reversal symmetry, and $\mathcal{Z}_2$ is a combination of twofold rotation in real space and chirality flip in orbital space. At the limit of $t_2/t=0$, $H_0$ acquires an even higher symmetry $D_3 \times SU(4) \times T \times \mathcal{Z}_2$.

\begin{figure}[h!]
\includegraphics[width=0.9\columnwidth]{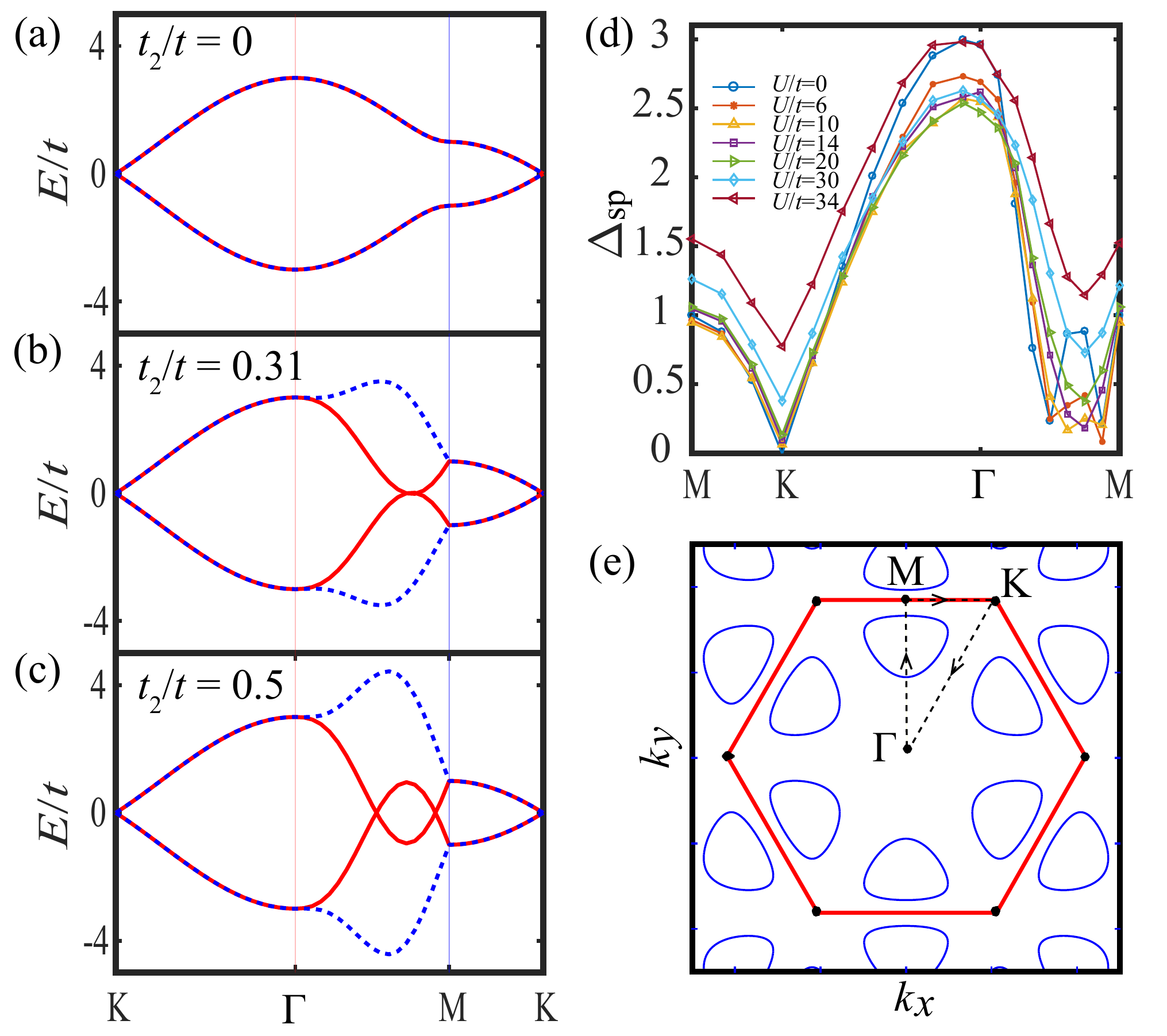}
\caption{(a)-(c) The band structure of the tight-binding model in Eq.~\ref{Seq:tb} for different $t_2/t$. Dirac points are at $\mathbf{K}$ and $\mathbf{K'}$, the crossing of bands within $\Gamma$--$M$ happens at $t_2/t=0.31$. (d) Single particle gap, $\Delta_{sp}$, along the high-symmetry path of BZ obtained from PQMC at $t_2/t = 0.5$ for $L=18$ system at different $U/t$. At $U=0$, due to the crossing within $\Gamma$-$M$, as shown in (c), $\Delta_{sp}$ is zero at $\mathbf{K}$, $\mathbf{K'}$ and the two crossing points. As $U/t$ increases, the crossing points are firstly gapped out at the $U$ value highlighted by the dashed line in Fig.1 of the main text, and further increasing of $U$ eventually gaps out the Dirac cones at $\mathbf{K}$ and $\mathbf{K'}$ at $U=U_c$. (e) The FS for the non-interacting system at $t_2/t=0.5$, with the red hexagon the first BZ. The pockets close to $\mathbf{M}$ points come from the encircling of dispersion within $\Gamma$-M.}
\label{Sfig:bandstructure}
\end{figure}

The band structure of $H_0$ is shown in Fig.~\ref{Sfig:bandstructure}. 
When $t_2/t = 0$, the orbital components are degenerate, so the band structure in Fig.\ref{Sfig:bandstructure}(a) is the same as that of the nearest hopping honeycomb lattice, i.e. graphene, with Dirac cones at momenta $\mathbf{K}$ and $\mathbf{K'}$ in BZ. Increasing $t_2/t$, upper and lower bands within $\Gamma$-M break the orbital degeneracy and the four bands manifest in this segment of high-symmetry path, with two inner ones move towards and eventually touch each other at $t_2/t\approx 0.31$, as shown in Fig.~\ref{Sfig:bandstructure} (b). Further increasing $t_2/t$ these two band crosses, as shown in Fig.\ref{Sfig:bandstructure} (c), and results in a uncommon Fermi surface (FS), which are the blue streched circles close to M points in two neighboring BZs, as shown in Fig.\ref{Sfig:bandstructure}(e). The parallel segments of these FS-s contribute high density of states to the tight-binding model.

Under the cluster charge interaction $U$, as shown in Fig.~\ref{Sfig:bandstructure} (d), the crossing points along $\Gamma$-M will be gapped out firstly, and consequently these FS-s will disappear and result in a jump in kinetic energy of the system, with the corresponding $U$ values hightlighted as the dash line in $t_2/t$--$U/t$ phase diagram in Fig.1 of the main text. After the jump, the FS-s will only contain the Dirac cones at $\mathbf{K}$ and $\mathbf{K'}$, and only when further increasing $U/t$, the Dirac cones will be gapped out as well through the $N=4$ Gross-Neveu chiral XY transition at $U_c$, as discussed in the main text.

\section{Projection QMC method and absence of sign-problem}

Since we are interested in the ground state properties of the system, the projection QMC is the method of choice~\cite{Assaad2008}.  In PQMC, one can obtain a ground state wave function $\vert \Psi_0 \rangle$ from projecting a trial wave function $\vert \Psi_T \rangle$  along the imaginary axis $\vert \Psi_0 \rangle = \lim\limits_{\Theta \to \infty} e^{-\frac{\Theta}{2} \mathbf{H}} \vert \Psi_T \rangle$, then observable can be calculated as
\begin{equation}
\label{eq:observablepqmc}
\langle \hat{O} \rangle = \frac{\langle \Psi_0 \vert \hat{O} \vert \Psi_0 \rangle}{\langle \Psi_0 \vert \Psi_0 \rangle} 
						= \lim\limits_{\Theta \to \infty} \frac{\langle \Psi_T \vert  e^{-\frac{\Theta}{2} \mathbf{H}} \hat{O}  e^{-\frac{\Theta}{2} \mathbf{H}} \vert \Psi_T \rangle}{\langle \Psi_T \vert  e^{-\Theta \mathbf{H}} \vert \Psi_T \rangle} .
\end{equation}

To evaluate overlaps in above equation, we performed Trotter decomposition and $\Theta$ is discretized into $L_\tau$ slices ($\Theta=L_\tau \Delta\tau$). Each slices $\Delta\tau$ is small and the systematic error is $\mathcal{O}(\Delta\tau^2)$. After the Trotter decomposition, we have
\begin{equation}
\langle\Psi_{T}|e^{-\Theta H}|\Psi_{T}\rangle=\langle\Psi_{T}|\left(e^{-\Delta\tau H_{U}}e^{-\Delta\tau H_{0}}\right)^{L_\tau}|\Psi_{T}\rangle+\mathcal{O}(\Delta{\tau}^{2})
\end{equation}
where the non-interacting and interacting parts of the Hamiltonian is separated. To treat the interacting part, one usually employ a Hubbard Stratonovich (HS) transformation to decouple the interacting quartic fermion term to fermion bilinears coupled to auxiliary fields. For the cluster charge interaction in the TBG model, we make use of a fourth order $SU(2)$ symmetric decoupling~\cite{Assaad2008,xu2018kekule}
\begin{equation}
e^{-\Delta\tau U(Q_{\varhexagon}-4)^{2}}=\frac{1}{4}\sum_{\{s_{\varhexagon}\}}\gamma(s_{\varhexagon})e^{\alpha\eta(s_{\varhexagon})\left(Q_{\varhexagon}-4\right)}
\label{eq:decompo}
\end{equation}
with $\alpha=\sqrt{-\Delta\tau U}$, $\gamma(\pm1)=1+\sqrt{6}/3$,
$\gamma(\pm2)=1-\sqrt{6}/3$, $\eta(\pm1)=\pm\sqrt{2(3-\sqrt{6})}$,
$\eta(\pm2)=\pm\sqrt{2(3+\sqrt{6})}$ and the sum is taken over the auxiliary fields $s_{\varhexagon}$ on each hexagon which can take four values $\pm2$ and $\pm1$. After tracing out the free fermions degrees of freedom, we obtain the following formula with a constant factor omitted
\begin{equation}
\langle\Psi_{T}|e^{-\Theta H}|\Psi_{T}\rangle=\sum_{\{s_{\varhexagon,\tau}\}}\left[\left(\prod_{\tau}\prod_{\varhexagon}\gamma(s_{\varhexagon,\tau})e^{-4\alpha\eta(s_{\varhexagon,\tau})}\right)\det\left[P^{\dagger}B(\Theta,0)P\right]\right]
\label{eq:mcweight}
\end{equation}
where $P$ is the coefficient matrix of trial wave function $|\Psi_T\rangle$. In the simulation,  we choose the ground state wavefunction of the half-filled non-interacting system (described by $H_0$) as the trial wave function. In the above formula, the $B$ matrix is defined as
\begin{equation}
B(\tau+1,\tau)=e^{V[\{s_{\varhexagon,\tau}\}]}e^{-\Delta_\tau K}
\end{equation}
and has properties $B(\tau_3,\tau_1)=B(\tau_3,\tau_2)B(\tau_2,\tau_1)$, where we have written the coefficient matrix of interaction part as $V[\{s_{\varhexagon,\tau}\}]$ and $K$ is the hopping matrix from the $H_0$.
The Monte Carlo sampling of auxiliary fields are further performed based on the weight defined in the sum of  Eq.~\eqref{eq:mcweight}. The measurements are performed near $\tau=\Theta/2$. Single particle observables are measured by Green's function directly and many body correlation functions are measured from the products of single-particle Green's function based on their corresponding form after Wick-decomposition. The equal time Green's function are calculated as
\begin{equation}
G(\tau,\tau)=1-R(\tau)\left(L(\tau)R(\tau)\right)^{-1}L(\tau)
\end{equation}
 with $R(\tau)=B(\tau,0)P$, $L(\tau)=P^{\dagger}B(\Theta,\tau)$. More technique details of PQMC method, please refer to Refs~\cite{blankenbecler1981monte,hirsch1985two,Assaad2008,xu2018kekule}. 

At half-filling, the model is sign-problem-free, this can be seen from the following analysis. Since the model is particle-hole symmetric at half-filling, one can perform a particle-hole transformation only for the orbital $l=2$, such transformation changes the cluster charge operator $Q_{\varhexagon} \equiv Q_{\varhexagon}^1 + Q_{\varhexagon}^2$ to $Q_{\varhexagon}' \equiv Q_{\varhexagon}^1 - Q_{\varhexagon}^2$ in the Hamiltonian. Here $Q_{\varhexagon}^{1}$/$Q_{\varhexagon}^{2}$ is cluster charge operator for orbital-1/orbital-2. Due to the specific form of Eq.~\ref{eq:decompo}, the fermion bilinears after HS transformation is invariant under the antiunitary transformation
\begin{equation}
\mathcal{U} = i\sigma_0 \tau_2 \mathcal{K},
\end{equation}
where $\sigma_0$ is identity in spin space, $\tau_2$ is second Pauli matrix in orbital space, $\mathcal{K}$ is complex conjugate. As all the fermion bilinears obeys above antiunitary symmetry, and $\mathcal{U}^2=-1$, these properties of our model and HS decoupling form give rise to a sufficient condition for sign-problem-free Monte Carlo simulations, as proved in Ref.~\cite{wu2005suff}.

\begin{figure}[h!]
\includegraphics[width=0.9\columnwidth]{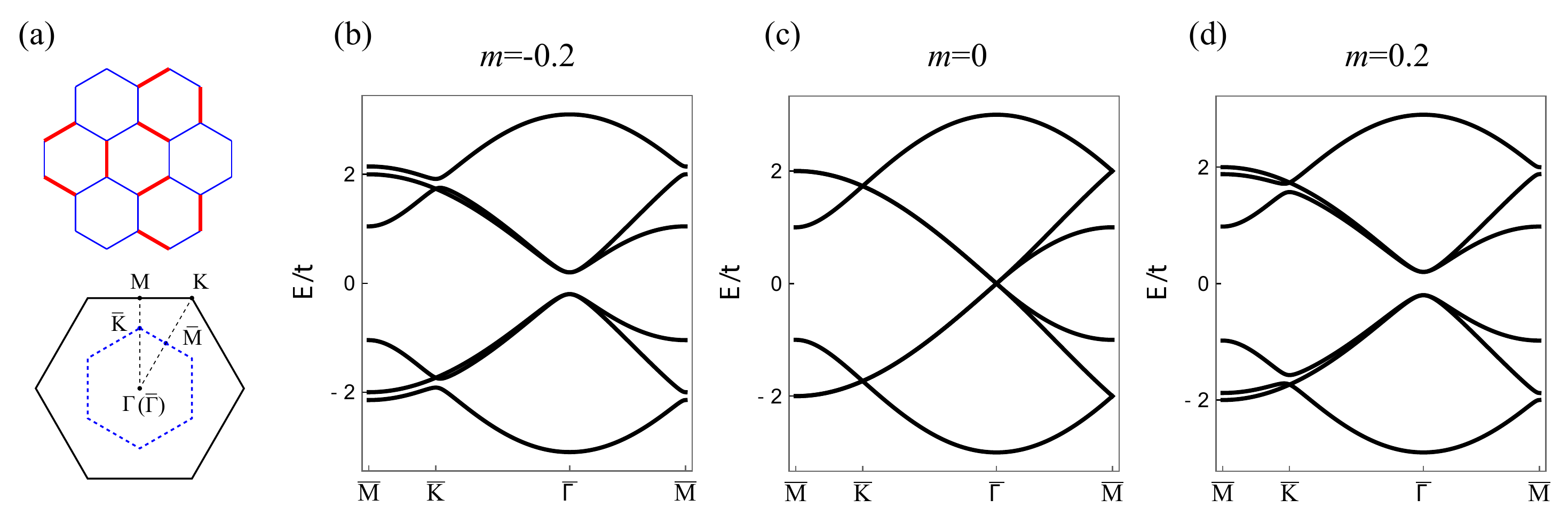}
\caption{(a) Bond modulation with $(1+\delta)t$ for red bonds and $(1-\delta)t$ for blue bonds. We define $m\equiv 2\delta$. The Brillouin zone is folded to smaller one (blue dash line) when $m \ne 0$. (b)-(d) Band structure with $m=-0.2$, $m=0$, and $m=0.2$.}
\label{Sfig:vbsmf}
\end{figure}

\section{$k \cdot p$ Hamiltonian for VBS phase}
In the VBS phase, we consider a mean field description with bond charge order, while forget the spin and orbital degrees of freedom. Then we can  define the
static VBS order as a modulation in the nearest neighbor hopping. As showed in the inset of Fig.1 of main text, 
there are two kinds of bonds and the hopping magnitude is defined as $(1+\delta)t$ and $(1-\delta)t$.
As such kind of bond modulation enlarges unit cell to $\sqrt{3}\times\sqrt{3}$, the tight binding Hamiltonian with enlarged unit cell writes
\begin{equation}
h(k)=-\tilde{t}\left(\begin{array}{cccccc}
0 & (1+m)e^{-ik_{y}} & 0 & e^{-i\frac{\sqrt{3}}{2}k_{x}+\frac{i}{2}k_{y}} & 0 & e^{i\frac{\sqrt{3}}{2}k_{x}+\frac{i}{2}k_{y}}\\
(1+m)e^{ik_{y}} & 0 & e^{i\frac{\sqrt{3}}{2}k_{x}-\frac{i}{2}k_{y}} & 0 & e^{-i\frac{\sqrt{3}}{2}k_{x}-\frac{i}{2}k_{y}} & 0\\
0 & e^{-i\frac{\sqrt{3}}{2}k_{x}+\frac{i}{2}k_{y}} & 0 & (1+m)e^{i\frac{\sqrt{3}}{2}k_{x}+\frac{i}{2}k_{y}} & 0 & e^{-ik_{y}}\\
e^{i\frac{\sqrt{3}}{2}k_{x}-\frac{i}{2}k_{y}} & 0 & (1+m)e^{-i\frac{\sqrt{3}}{2}k_{x}-\frac{i}{2}k_{y}} & 0 & e^{ik_{y}} & 0\\
0 & e^{i\frac{\sqrt{3}}{2}k_{x}+\frac{i}{2}k_{y}} & 0 & e^{-ik_{y}} & 0 & (1+m)e^{-i\frac{\sqrt{3}}{2}k_{x}+\frac{i}{2}k_{y}}\\
e^{-i\frac{\sqrt{3}}{2}k_{x}-\frac{i}{2}k_{y}} & 0 & e^{ik_{y}} & 0 & (1+m)e^{i\frac{\sqrt{3}}{2}k_{x}-\frac{i}{2}k_{y}} & 0
\end{array}\right)
\end{equation}
where $m\equiv 2\delta$, $\tilde{t}=(1-\delta)t$.
Fig.~\ref{Sfig:vbsmf} shows band structure with different mass term $m$. We can see the bond modulation folds Dirac point K and K' of original BZ to $\Gamma$ point in VBS BZ (small BZ) and open a gap when $m\ne 0$ and the gap is exactly $m$. This can be seen more explicitly if we expand above tight binding Hamiltonian around $\Gamma$ point and perform a further down folding from six bands to four bands with unitary transformation
\begin{equation}
U=\left(\begin{array}{ccc}
-1 & -1 & 1\\
0 & 1 & 1\\
1 & 0 & 1
\end{array}\right)\otimes \tau^{0}.
\end{equation}
Then we get  a $k \cdot p$ Hamiltonian 
$H_{\text{eff}}(\vec{k}) = -t\left( \tilde{\vec{k}}\cdot \tilde{ \vec{s}}\tau^2 + ms^0 \tau^1 \right)$
where only linear terms of $k$ and $m$ are reserved.
We change the basis a little bit ($\tau^1 \rightarrow \tau^3$ and $\tau^2 \rightarrow \tau^2$), and get the final form as shown in Eq.(4) of the main text.  For convenient, we also show it here
\begin{equation}
\label{Seq:kp}
H_{\text{eff}}(\vec{k}) = -t\left( \tilde{\vec{k}}\cdot \tilde{ \vec{s}}\tau^2 + ms^0 \tau^3 \right),
\end{equation}
where momentum $\tilde{\vec{k}}\equiv (\frac 3 2 k_y, \frac {\sqrt{3}} {2} ik_x, \sqrt{3}k_x)$, the vector $\tilde{\vec{s}}=(s^1, s^2, s^3)$. Here $s^i$ and $\tau^i$ are Pauli matrices in two different spaces. From the $k\cdot p$ Hamiltonian, it is clear that the bond modulation plays the role of a mass term, and there is a sign change in it across the pVBS-cVBS transition.

\section{Critical exponents for 3D $N=4$ Gross-Nevue chiral XY universality class}
We compare the critical exponents for 3D $N=4$ Gross-Nevue chiral XY universality class calculated by different methods. As showed in Table~\ref{Stable:exponents}, the critical exponents we get are comparable with existed results. We want to remark, there are still differences between the critical exponents got from different methods. Much more efforts in the direction of developments of advanced analytical techniques for higher order calculations, as well as the developments of technically sound numeric methods to achieve much larger system sizes are requiring.
\begin{table}[ht]
    \caption{Comparison of critcal exponents for 3D $N=4$ Gross-Nevue chiral XY universality class calculated by different methods.} % title of Table
    \centering % used for centering table
    %\bgroup
    \def\arraystretch{1.5}
    \begin{tabular}{c  c  c } % centered columns (4 columns)
    \hline\hline %inserts double horizontal lines
    $N=4$ & $\eta$ & $\nu$ \\ [0.0ex] % inserts table
    %heading
    \hline % inserts single horizontal line
     Monte Carlo, \it this work   & \ \  0.80(2)  &  \ \ 1.01(3)   \\ [0.0ex] % [1ex] adds vertical space
    \hline
     $4-\epsilon$, four loop, $P_{[2/2]}$~\cite{zerf2017four}  & \ \  0.929  & \ \  1.130   \\ [0.0ex] 
    \hline
    $4-\epsilon$, four loop, $P_{[3/1]}$~\cite{zerf2017four}  & \ \  0.911  & \ \  1.130   \\ [0.0ex]
    \hline
    functional RG (LPA')~\cite{classen2017fluctuation} & \ \  0.946 & \ \ 1.082 \\ [0.0ex]
    \hline %inserts single line
    $4-\epsilon$, two loop~\cite{rosenstein1993critical}  &  \ \ 0.82  &  \ \ 0.97   \\ [0.0ex] 
    \hline                      
     Monte Carlo~\cite{li2017fermion} & \ \  0.80(4)  &  \ \ 1.11(3)   \\ [0.0ex]
    \hline
     large-$N$~\cite{hands1993} & \ \ 0.932 & \ \ 1.135 \\ [0.0ex]
    \hline                      
    %\multicolumn{3}{l}{\textsuperscript{*}\footnotesize{This work.}}
    \end{tabular}
    \label{Stable:exponents} % is used to refer this table in the text
\end{table}

\end{document}